\documentstyle[epsf,psfig]{article}

\catcode`\@=11
\long\def\@makefntext#1{
\protect\noindent \hbox to 3.2pt {\hskip-.9pt
$^{{\eightrm\@thefnmark}}$\hfil}#1\hfill}               

\def\@makefnmark{\hbox to 0pt{$^{\@thefnmark}$\hss}}    

\def\ps@myheadings{\let\@mkboth\@gobbletwo
\def\@oddhead{\hbox{}
\rightmark\hfil\eightrm\thepage}
\def\@oddfoot{}\def\@evenhead{\eightrm\thepage\hfil
\leftmark\hbox{}}\def\@evenfoot{}
\def\sectionmark##1{}\def\subsectionmark##1{}}

\oddsidemargin=\evensidemargin
\newcounter{sectionc}\newcounter{subsectionc}\newcounter{subsubsectionc}
\renewcommand{\section}[1] {\vspace{12pt}\addtocounter{sectionc}{1}
\setcounter{subsectionc}{0}\setcounter{subsubsectionc}{0}\noindent
        {\tenbf\thesectionc. #1}\par\vspace{5pt}}
\renewcommand{\subsection}[1] {\vspace{12pt}\addtocounter{subsectionc}{1}
        \setcounter{subsubsectionc}{0}\noindent
        {\bf\thesectionc.\thesubsectionc. {\kern1pt \bfit #1}}\par\vspace{5pt}}
\renewcommand{\subsubsection}[1] {\vspace{12pt}\addtocounter{subsubsectionc}{1}
        \noindent{\tenrm\thesectionc.\thesubsectionc.\thesubsubsectionc.
        {\kern1pt \tenit #1}}\par\vspace{5pt}}
\newcommand{\nonumsection}[1] {\vspace{12pt}\noindent{\tenbf #1}
        \par\vspace{5pt}}

\newcounter{appendixc}
\newcounter{subappendixc}[appendixc]
\newcounter{subsubappendixc}[subappendixc]
\renewcommand{\thesubappendixc}{\Alph{appendixc}.\arabic{subappendixc}}
\renewcommand{\thesubsubappendixc}
        {\Alph{appendixc}.\arabic{subappendixc}.\arabic{subsubappendixc}}

\renewcommand{\appendix}[1] {\vspace{12pt}
        \refstepcounter{appendixc}
        \setcounter{figure}{0}
        \setcounter{table}{0}
        \setcounter{lemma}{0}
        \setcounter{theorem}{0}
        \setcounter{corollary}{0}
        \setcounter{definition}{0}
        \setcounter{equation}{0}
        \renewcommand{\thefigure}{\Alph{appendixc}.\arabic{figure}}
        \renewcommand{\thetable}{\Alph{appendixc}.\arabic{table}}
        \renewcommand{\theappendixc}{\Alph{appendixc}}
        \renewcommand{\thelemma}{\Alph{appendixc}.\arabic{lemma}}
        \renewcommand{\thetheorem}{\Alph{appendixc}.\arabic{theorem}}
        \renewcommand{\thedefinition}{\Alph{appendixc}.\arabic{definition}}
        \renewcommand{\thecorollary}{\Alph{appendixc}.\arabic{corollary}}
        \renewcommand{\theequation}{\Alph{appendixc}.\arabic{equation}}
        \noindent{\tenbf Appendix \theappendixc #1}\par\vspace{5pt}}
\newcommand{\subappendix}[1] {\vspace{12pt}
        \refstepcounter{subappendixc}
        \noindent{\bf Appendix \thesubappendixc. {\kern1pt \bfit #1}}
        \par\vspace{5pt}}
\newcommand{\subsubappendix}[1] {\vspace{12pt}
        \refstepcounter{subsubappendixc}
        \noindent{\rm Appendix \thesubsubappendixc. {\kern1pt \tenit #1}}
        \par\vspace{5pt}}

\newcommand{\textlineskip}{\baselineskip=13pt}
\newcommand{\smalllineskip}{\baselineskip=10pt}

\def\eightcirc{
\begin{picture}(0,0)
\put(4.4,1.8){\circle{6.5}}
\end{picture}}
\def\eightcopyright{\eightcirc\kern2.7pt\hbox{\eightrm c}}

\newcommand{\copyrightheading}[1]
        {\vspace*{-2.5cm}\smalllineskip{\flushleft
        {\footnotesize International Journal of Modern Physics B, #1}\\
        {\footnotesize $\eightcopyright$\, World Scientific Publishing
         Company}\\
         }}

\newcommand{\publisher}[2]{{\begin{center}\footnotesize\smalllineskip
        Received #1\\
        Revised #2
        \end{center}
        }}

\def\abstracts#1#2#3{{
        \centering{\begin{minipage}{4.5in}\baselineskip=10pt\footnotesize
        \parindent=0pt #1\par
        \parindent=15pt #2\par
        \parindent=15pt #3
        \end{minipage}}\par}}

\renewenvironment{thebibliography}[1]                   
        {\frenchspacing
         \ninerm\baselineskip=11pt
         \begin{list}{\arabic{enumi}.}
        {\usecounter{enumi}\setlength{\parsep}{0pt}
         \setlength{\leftmargin 12.7pt}{\rightmargin 0pt} 
         \setlength{\itemsep}{0pt} \settowidth
        {\labelwidth}{#1.}\sloppy}}{\end{list}}

\newcounter{itemlistc}
\newcounter{romanlistc}
\newcounter{alphlistc}
\newcounter{arabiclistc}

\newcommand{\fcaption}[1]{
        \refstepcounter{figure}
        \setbox\@tempboxa = \hbox{\footnotesize Fig.~\thefigure. #1}
        \ifdim \wd\@tempboxa > 5in
           {\begin{center}
        \parbox{5in}{\footnotesize\smalllineskip Fig.~\thefigure. #1}
            \end{center}}
        \else
             {\begin{center}
             {\footnotesize Fig.~\thefigure. #1}
              \end{center}}
        \fi}

\newcommand{\tcaption}[1]{
        \refstepcounter{table}
        \setbox\@tempboxa = \hbox{\footnotesize Table~\thetable. #1}
        \ifdim \wd\@tempboxa > 5in
           {\begin{center}
        \parbox{5in}{\footnotesize\smalllineskip Table~\thetable. #1}
            \end{center}}
        \else
             {\begin{center}
             {\footnotesize Table~\thetable. #1}
              \end{center}}
        \fi}

\def\@citex[#1]#2{\if@filesw\immediate\write\@auxout
        {\string\citation{#2}}\fi
\def\@citea{}\@cite{\@for\@citeb:=#2\do
        {\@citea\def\@citea{,}\@ifundefined
        {b@\@citeb}{{\bf ?}\@warning
        {Citation `\@citeb' on page \thepage \space undefined}}
        {\csname b@\@citeb\endcsname}}}{#1}}

\newif\if@cghi
\def\cite{\@cghitrue\@ifnextchar [{\@tempswatrue
        \@citex}{\@tempswafalse\@citex[]}}
\def\citelow{\@cghifalse\@ifnextchar [{\@tempswatrue
        \@citex}{\@tempswafalse\@citex[]}}
\def\@cite#1#2{{$\null^{#1}$\if@tempswa\typeout
        {IJCGA warning: optional citation argument
        ignored: `#2'} \fi}}

\def\pmb#1{\setbox0=\hbox{#1}
        \kern-.025em\copy0\kern-\wd0
        \kern.05em\copy0\kern-\wd0
        \kern-.025em\raise.0433em\box0}

\def\fnt#1#2{\footnotetext{\kern-.3em
        {$^{\mbox{\scriptsize #1}}$}{#2}}}

\def\fpage#1{\begingroup
\voffset=.3in
\thispagestyle{empty}\begin{table}[b]\centerline{\footnotesize #1}
        \end{table}\endgroup}

\font\tenrm=cmr10
\font\tenit=cmti10
\font\tenbf=cmbx10
\font\bfit=cmbxti10 at 10pt
\font\ninerm=cmr9
\font\nineit=cmti9
\font\ninebf=cmbx9
\font\eightrm=cmr8

\newcommand{\bfm}[1]{\mbox{\boldmath $#1$}}
\newcommand{\Eq}[1]{Eq.(#1)}

\newcommand{\la}{\langle}
\newcommand{\ra}{\rangle}
\newcommand{\rta}{\rightarrow}

\newcommand{\nIs}{n_{1\s}}
\newcommand{\nJs}{n_{2\s}}

\newcommand{\R}{\Rv}
\newcommand{\Rv}{\bfm{R}}
\newcommand{\rv}{\bfm{r}}

\newcommand{\RIo}{\Rv_1^0}
\newcommand{\RJo}{\Rv_2^0}
\newcommand{\s}{\sigma}

\newcommand{\uv}{\bfm{u}}

\newcommand{\uI}{\uv^{~}_1}
\newcommand{\uJ}{\uv^{~}_2}

\newcommand{\beq}{\begin{equation}}
\newcommand{\eeq}{\end{equation}}

\def\qed{\hbox{${\vcenter{\vbox{                        
   \hrule height 0.4pt\hbox{\vrule width 0.4pt height 6pt
   \kern5pt\vrule width 0.4pt}\hrule height 0.4pt}}}$}}

\def\bsc{{\sc a\kern-6.4pt\sc a\kern-6.4pt\sc a}}       
\def\bflatex{\bf L\kern-.30em\raise.3ex\hbox{\bsc}\kern-.14em
T\kern-.1667em\lower.7ex\hbox{E}\kern-.125em X}

\begin{document}


\normalsize\textlineskip
\thispagestyle{empty}
\setcounter{page}{1}

\copyrightheading{}                     

\vspace*{0.88truein}

\fpage{1}
\centerline{\bf MODEL CALCULATION OF ELECTRON-PHONON COUPLINGS}
\vspace*{0.035truein}
\centerline{\bf IN A DIMER WITH A NON-DEGENERATE ORBITAL}
\vspace*{0.37truein}
\centerline{M. ACQUARONE}
\vspace*{0.015truein}
\centerline{\it C.N.R.-G.N.S.M., Unita' I.N.F.M.\ di Parma, Dipartimento di
Fisica}
\baselineskip=10pt
\centerline{\it Universita' di Parma, Parma, I-43100,
Italy}
\vspace*{10pt}
\centerline{\normalsize and}
\vspace*{10pt}
\centerline{C. NOCE}
\vspace*{0.015truein}
\centerline{\it Unit\`a I.N.F.M.,
Dipartimento di Scienze Fisiche ''E.R. Caianiello''}
\baselineskip=10pt
\centerline{\it Universit\`a di Salerno, Baronissi, I-84081, Italy}
\vspace*{0.225truein}
\publisher{( )}{( )}

\vspace*{0.21truein}
\abstracts{ We
evaluate all the  electron-phonon couplings derived from the one-body
and two-body electronic interactions,
in both the adiabatic and extreme non-adiabatic
limit, for a dimer with a non-degenerate orbital built
from atomic wave functions of Gaussian shape. We find largely different values
 of the coupling parameters in the two limits, as well as
different expressions of the corresponding terms in the
Hamiltonian. Depending on the distance between the dimer ions,
some of the two-body couplings are comparable, or even larger than
the one-body ones.  } {}{}



\vspace*{1pt}\textlineskip      
\section{Introduction}  
\vspace*{-0.5pt}
\noindent
The models containing interactions of electrons and phonons have been
the object of intense study in the recent years, mainly due to the
discovery of high $T_c$ superconductors ($HTS$) and colossal
magnetoresistance manganites ($CMR$). When trying to apply the
theoretical results to the real materials,
 however,  one is confronted with the
difficulty that a large degree of arbitrariness is
present, stemming from two origins. First,
 different forms of interactions
have been introduced, essentially on an {\it ad hoc} basis.  Second, the
value of the interaction parameters are estimated from the
experimental data, which, however, can be interpreted only after a
definite theoretical model, i.e. a set of electron-phonon interactions,
 has been assumed to be relevant for the material under measurement.
 Therefore, besides unavoidable
quantitative approximations in fitting the  data,
 also  qualitative uncertainties are introduced {\it a priori} in the
analysis.
  The purpose of the present work is to
provide, in a simple but not trivial model, both
  rigorous prescriptions about the admissible electron-phonon
 interaction terms, and the quantitative evaluation of the interaction
strength parameters. The model we consider is a dimer where the
electrons occupy a non-degenerate orbital described by Wannier
functions built from atomic orbitals of Gaussian shape.  All the
one- and two-body electronic parameters in that model have been
analytically evaluated in Ref.1. The electron-phonon interactions
will be evaluated following the general definitions introduced in
Refs.2 and 3.\par The structure of the paper is as follows: in
Sec.2 we describe the model and the general procedure of
evaluation; Secs.3 and 4 are devoted to the analytical evaluation
of the coupling parameters in the non-adiabatic, and adiabatic
limits, respectively, also defining the correct form of the
corresponding terms in the Hamiltonian; Sec.5 discusses the
numerical evaluation of the coupling parameters; Sec.6 is devoted
to the conclusions. The Appendices contain details of the
analytical calculations.

\section{The Model}

\noindent In a general electron-phonon Hamiltonian
$H=H_{el}+H_{ph}+H_{el-ph}  \label{ham1}$ the interacting term
$H_{el-ph}$ originates from the  development of the
electronic part $H_{el}$ to first order in the phonon-induced  displacement of
the ion positions. In our analysis we shall consider  the terms due to
 the
development of both the one-body and two-body contributions
to the electronic Hamiltonian, which, in standard
notation{\cite{acquarone1}, reads:
\[
H_{el}=\epsilon \sum_{\sigma }(n_{1\sigma }+n_{2\sigma})+\sum_{\sigma}
[t+X(n_{1-\sigma }+n_{2-\sigma})](c_{1\sigma }^{\dagger }c_{2\sigma }+H.c.)
+U(n_{1\uparrow}n_{1\downarrow}+n_{2\uparrow}n_{2\downarrow})
\]
\begin{equation}
+Vn_1n_2-2J_zS^z_1S^z_2-J_{xy}(S^+_1S^-_2+H.c.)
+P(c^{\dagger}_{1\uparrow}c^{\dagger}_{1\downarrow}
c^{}_{2\downarrow}c^{}_{2\uparrow}+H.c.).
 \label{helbare}
\end{equation}
The electrons are assumed to occupy a non-degenerate orbital described
by Wannier functions built from atomic orbitals of Gaussias shape.

\subsection{Undisplaced Wannier functions}

\noindent The ions occupy the undeformed positions $\RIo=(-a/2,0,0)$ and $
\RJo=(a/2,0,0)
$. The unit vector ${\bfm e}_{12}=(\RJo-\RIo)/a=(1,0,0)$
points from ion 1 towards ion 2;
if not specified otherwise, $\bfm{e}=\bfm{e}_{12}$.
The position of the electron is $\rv=(x,y,z)$. We associate to
each site $i=1,2$ a
Gaussian atomic-like orbital $\phi_i(\rv-\R_i)$. By defining
$N\equiv\left (2/ \pi\right)^{3/4}\Gamma^{3/2}$, so that
$<\phi_i|\phi_i>=1$, they read:
\[
\phi_1(\rv-\RIo)=
N\exp\left\{-\Gamma^2\left[\left(x+a/2\right)^2+y^2+z^2\right]\right\},
\]
\beq
\phi_2(\rv-\RJo)=
N\exp\left\{-\Gamma^2\left[\left(x-a/2\right)^2+y^2+z^2\right]\right\}.
\label{wan2}
\eeq
Their overlap $S_0 \equiv \la \phi_1|\phi_2\ra =
\exp(-\Gamma^2a^2/2)$ is non-vanishing. Then the Wannier functions
$\Psi_1, \Psi_2$ can be written as:
\[
\Psi_1(\rv -\RIo,\rv- \RJo)=A(S_0) \phi_1(\rv-\RIo) + B(S_0)
\phi_2(\rv-\RJo),
\]
\beq
\Psi_2(\rv-\RIo,\rv- \RJo)=B(S_0) \phi_1(\rv-\RIo) + A(S_0)
\phi_2(\rv-\RJo).
\label{wan1}
\eeq
The requirement
$<\Psi_i|\Psi_j>=\delta_{ij}$ yields the mixing coefficients $A,B$
as: \beq A(S_0)\equiv {1 \over 2} \left [ {{1}\over{ \sqrt{1+S_0}
}} +
 {{1}\over{ \sqrt{1-S_0} }} \right],
\qquad
B(S_0)\equiv {1 \over 2} \left [ {{1}\over{ \sqrt{1+S_0} }} -
 {{1}\over{ \sqrt{1-S_0} }} \right].
\label{wan4}
\eeq
In the limit $S_0\rightarrow 0$, one has $\Psi_i \rightarrow \phi_i$
$(i=1,2)$,
so that one can connect unambiguously each Wannier function to a well
defined site. \par

\subsection{Displaced Wannier functions}

\noindent We shall consider only deformations $\uI=u_1\bfm{e}_{12}$ and
$\uJ=u_2\bfm{e}_{12}$ altering the length of the dimer,
 so that  only their $x$-components are non vanishing.
Their signs are defined, on both sites, with
respect to the same unit vector $\bfm{e}_{12}$, so as
 to have a single reference system. Symmetry requirements impose
$\uI=-\uJ$ so that the relative displacement $u\equiv
u_2-u_1=-2u_1=2u_2$. Even though we shall never introduce the
exact time dependency in the following calculation, it is
important to keep in mind that the displacements vary periodically
in time with the frequency $\Omega$ of the phonon. The
instantaneous positions of the ions are therefore defined as
\beq
\R_1=\RIo+\uI=(-a/2+u_1, 0,0),\qquad \R_2=\RJo+\uJ=(a/2+u_2, 0,0).
\label{istpos}
\eeq
\noindent In the adiabatic limit, the orbitals
on each site are assumed to
 adjust instantaneously to the displaced  positions of the ions.
 Then, if we define
\[
\phi_1[\rv-(\RIo+\bfm{u}_1)]=
N\exp\left\{-\Gamma^2\left[\left(x+a/2-u_1\right)^2+y^2+z^2\right]\right\},
\]
\beq
\phi_2[\rv-(\RJo+\bfm{u}_2)]=
N\exp\left\{-\Gamma^2\left[\left(x-a/2-u_2\right)^2+y^2+z^2\right]\right\},
\label{wan2d}
\eeq
the Wannier functions $\Psi_i$ ($i=1,2$)are
written as:
\[
\Psi_1(\RIo,\RJo,u_1,u_2)=A(u_1,u_2) \phi_1(\rv-\RIo-\bfm{u}_1) +
B(u_1,u_2) \phi_2(\rv-\RJo-\bfm{u}_2),
\]
\beq
\Psi_2(\RIo,\RJo,u_1,u_2)=B(u_1,u_2)
\phi_1(\rv-\RIo-\bfm{u}_1) + A(u_1,u_2)
\phi_2(\rv-\RJo-\bfm{u}_2). \label{wan1d}
\eeq
In principle we
have four coefficients $A(u_1,u_2),B(u_1,u_2)$ to determine. We
make the reasonable guess that, as for $u_i\rta 0$ the $A$'s ( and
the $B$'s) have the same shape on both sites, then they keep the
same shape also for $u_i \ne 0$.  We shall define $A_0, B_0,S_0$
as $\lim_{u_1,u_2\rta 0}A(u_1,u_2)$ etc. As $\la
\phi_1(u_1)|\phi_1(u_1)\ra=\la \phi_2(u_2)|\phi_2(u_2)\ra=1$ still
holds, by defining
 $S\equiv \exp\{-( {{\Gamma^2}/{2}})[a-(u_2-u_1)]^2\}  $
the  condition $\la \Psi_i(u_1,u_2)|\Psi_j(u_1,u_2)\ra=\delta_{ij}$
 now yields:
\beq
A(S)\equiv {1 \over 2} \left [ {{1}\over{ \sqrt{1+S} }} +
 {{1}\over{ \sqrt{1-S} }} \right],
\qquad
B(S)\equiv {1 \over 2} \left [ {{1}\over{ \sqrt{1+S} }} -
 {{1}\over{ \sqrt{1-S} }} \right].
\label{AsBs}
\eeq
Namely, $A(S)$ and $B(S)$ depend on $S(u)$ as $A_0$ and $B_0$ depend
on $S_0$.\par
\subsection{The electronic parameters}

\noindent Given the Wannier functions, all the electronic
parameters can be evaluated: this was done in Ref.1.
 To make clear our method of calculation, it is
convenient to explicitate
 the one-body electronic interactions,
corresponding to the local energy $\epsilon$
and to the hopping amplitude $t$.

\beq
\epsilon_i=\int\Psi_i^{\ast}(\rv,\R_1,\R_2)
\left[-{{\hbar^2}\over{2m}}\nabla^2+V_1(\rv-\R_1)+V_2(\rv-\R_2)\right]
\Psi_i^{~}(\rv,\R_1,\R_2) d^3\rv,
\label{epsilon0}
\eeq
\beq
t=\int \Psi_1^{\ast}(\rv,\R_1,\R_2)
\left[-{{\hbar^2}\over{2m}}\nabla^2+V_1(\rv-\R_1)+V_2(\rv-\R_2)\right]
\Psi_2^{~}(\rv,\R_1,\R_2) d^3\rv.
\label{hopping}
\eeq
In the
formulas above, the potentials originating from the ion cores at
the displaced positions $\R_1$ and $\R_2$ are:
\[
V_1\equiv V(\rv-\R_1)=-e^2Z \left[\left(x+{a \over 2}-u_1\right)^2
+y^2+z^2\right]^{-1/2},
\]
\beq
V_2\equiv V(\rv-\R_2)=-e^2Z\left[\left(x-{a \over 2}- u_2
\right)^2+y^2+z^2\right]^{-1/2}, \label{hol5}
\eeq
where $-e$ is
the electron charge, and $+Ze$ is the charge of the ion core. \par
The local energy  $\epsilon$ (actually site-independent)
 can be decomposed into three terms,
respectively corresponding to the contributions from the Laplacian
kinetic operator ($\epsilon_{\nabla}^{~}$) and from each one of
the  ionic potentials ($\epsilon_{V_1}^{~},\epsilon_{V_2}^{~}$).
The modulation of the term where both charge distribution and
potential refer to the same ion gives rise to an Holstein-type
coupling. The other term, where the charge around one ions feels
the displaced potential of the other ion, we shall call
"crystal-field" coupling. In the literature\cite{barisic}
 this latter term is usually assumed to be negligible
but we shall see that actually is of the same order as the other
ones, and can even become the dominant one.\par The modulation of
the hopping amplitude $t$ gives rise to the so-called
Su-Schrieffer-Heeger ($SSH$) interaction\cite{SSH}. Similarly to
$\epsilon$, it can be decomposed into  kinetic ($t^{~}_{\nabla}$)
 and  potential ($t{~}_{V_1,V_2}$) contributions.  \par
We need to  distinguish between the adiabatic (labelled {\it ad})
 and the non-adiabatic  (labelled {\it na})
limit in evaluating the electron-phonon interactions, because the
integrals have different kernels in the two cases. Indeed, in the
adiabatic limit, when $\hbar\Omega<t$,
the  displacements affect both the potentials and
the electronic Wannier functions, expressing the requisite that the
electronic charge distribution adjusts itself instantaneously at the
position of the ions. \par
We shall schematize the opposite situation $\hbar\Omega>t$,
where the electrons are
slower than the ions, as realized by  the
electronic charge distribution staying centred around the undisplaced
ion position, while the potentials are centred on the displaced ions.
 We shall call this the extreme anti-adiabatic limit. Not only the
strength of the interactions, but
also the form of the terms contributing to the Hamiltonian will turn out
to differ in the two cases.\par
 The anti-adiabatic limit will be treated first, as the notation is easier
 to establish in that case.
\vfill\eject
\section{Couplings in the non-adiabatic limit}

 If the Wannier functions keep being centred around the
undisplaced ionic positions, neither $\epsilon^{~}_{\nabla}$ nor
$t^{~}_{\nabla}$
 does change, therefore
 no electron-phonon coupling originates from them.
The couplings derived from the two-body interactions are also identically
vanishing in this limit, because they involve the Wannier
functions and the inter-electronic Coulomb potential
 which are both insensitive to the displacements of the ions.\par
 The only non-vanishing
types of electron-phonon non-adiabatic couplings  are those arising
from the variation of  the potential contributions to $\epsilon$ and
$t$.

\subsection{The Holstein-type coupling $g_0$}

\noindent The Holstein-type coupling $g_0$ is usually introduced
as the site-independent amplitude of a term in the Hamiltonian
connecting the local charge with the local deformation:
\beq
g_0\sum_\s (\nIs u_1+\nJs u_2). \label{g0def1}
\eeq
\noindent This
interaction originates from the perturbation of the on-site atomic
energy $\epsilon_{0}$ due to the ion motion in the non-adiabatic
limit, namely:
\[
\langle\Psi_i(\RIo,\RJo)|V_i(\R_i)|\Psi_i(\RIo,\RJo)\rangle-
\langle\Psi_i(\RIo,\RJo)|V_i(\R_{i}^0)|\Psi_i(\RIo,\RJo)\rangle
\]
\beq \equiv g_0^{(i)}n_iu_i +{\cal{O}}(u_i^2),\qquad (i=1,2)
\label{g0def} \eeq
 Our $g_0^{(i)}$
is therefore different from the interaction defined by
Holstein\cite{holstein}
because he considered the interaction as due to the
variation of the internuclear distance between the two components of
each dimer attached at the nodes of a frozen chain. We use the
"Holstein" label  because this interaction  has
the same form as the traditional Holstein term.
We also allow for a possible site-dependency of $g_0^{(i)}$, to show
later that this is not the case.\par
For the ion at $\R_1$ one writes  $g_0^{(1)} $ as:
\beq
g_0^{(1)} \equiv
\left(\lim_{u_1\rta 0}{{\partial\epsilon^{(1)}_{V_1}}\over{\partial u_1}}\right)
=\left(\lim_{u_1\rta 0} {{\partial } \over{\partial u_1}}\right)
 \int_{-\infty}^{\infty}\Psi_1(x,y,z,\RIo,\RJo)^2V_1(u_1^{~})dxdydz.
\label{g01}
\eeq
In the integral
\[
 \int_{-\infty}^{\infty}\Psi_1(x,y,z,\RIo,\RJo)^2V_1(u_1^{~})dxdydz =
\]
\begin{equation}
\int^{\infty}_{-\infty}
\left\{A\phi_1\left[\left(x+{{a}\over{2}}\right),y,z\right]+
B\phi_2\left[\left(x-{{a}\over{2}}\right),y,z\right]\right\}^2
V_1\left[\left(x+{{a}\over{2}}-u_1^{~}\right),y,z\right]dxdydz,
\label{g0def2}
\end{equation}
let us change variable form $x$ to $p=x+a/2-u_1^{~}$, to shift the
origin of $x$ coordinate onto the displaced ion, without changing
the shape of the Wannier functions. The reason is to avoid the
derivation with respect to $u_1^{~}$ of $V_1(u_1^{~})$ which is
discontinuous in the integration range.
 The above expression changes into:
\begin{equation}
\int^{\infty}_{-\infty}
\left\{A\phi_1\left[\left(p+u_1^{~}\right),y,z\right]+
B\phi_2\left[\left(p-a+u_1^{~}\right),y,z\right]\right\}^2
V_1\left[(p,y,z)\right]dxdydz.
\label{g0u}
\end{equation}
In this form  the kernel has a continuous derivative with respect to $u_1^{~}$
everywhere. Notice also that $A$ and $B$ still do not depend on $u_1^{~}$.
To derive with respect to $u_1^{~}$ Eq.\ref{g0u}, we use
$u_2^{~}=-u_1^{~}$ and notice that:
\begin{equation}
\lim_{u_1^{~}\rta
0}{{\partial\phi_1^{~}(p+u_1^{~},y,z)}\over{\partial u_1^{~}}}
={{\partial\phi_1^{~}(x,y,z)}\over{\partial x}}, \qquad
\lim_{u_2^{~}\rta 0}
{{\partial\phi_2^{~}(p-a-u_2^{~},y,z)}\over{\partial u_2^{~}}}
=-{{\partial\phi_2^{~}(x,y,z)}\over{\partial x}}, \label{derphi}
\end{equation}
yielding:
\[
\lim_{u_1^{~}\rta 0}{{\partial}\over{u_1^{~}}}
\left\{A\phi_1\left[\left(p+u_1^{~}\right),y,z\right]+
B\phi_2\left[\left(p-a+u_1^{~}\right),y,z\right]\right\}^2
\]
\begin{equation}
=2\Psi_1^{~}\left[A{{\partial\phi_1^{~}}\over{\partial x}}
+B{{\partial\phi_2^{~}}\over{\partial x}}\right]
=2\Psi_1^{~}{{\partial\Psi_1^{~}}\over{\partial x}}.
\label{kernel1}
\end{equation}
Therefore we can write:
\begin{equation}
g_0^{(1)}=2\int^{\infty}_{-\infty}
\Psi_1^{~}{{\partial\Psi_1^{~}}\over{\partial x}} V_1^{~}dxdydz.
\label{hol4}
\end{equation}
By substituting $\Psi_1$  and its derivative
into Eq.{\ref{hol4}} and  by noticing that
 $\la \phi_1|\left(x+{a / 2} \right)V_1|\phi_1\ra =0$,
as the kernel is odd in $x+a/2$,  we obtain:
\beq g_0^{(1)}=-4\Gamma^2\int^{\infty}_{-\infty} \left(B^2\phi_2^2
+2AB\phi_1\phi_2\right)xV_1 dxdydz +
2\Gamma^2aB^2\int^{\infty}_{-\infty}\phi_2^2 V_1dxdydz.
\label{hol6} \eeq The integrals in Eq.{\ref{hol6}} are listed in
Appendix 2. The final result is: \beq
g_0^{(1)}={{2\Gamma\sqrt{2/\pi} } \over {a}} \left\{B^2
\left[F_0(2a^2\Gamma^2)  -S^4\right] +4ABS \left[
F_0\left({{a^2\Gamma^2}\over{2}}\right)-S \right] \right\}.
\label{hol27c} \eeq
\noindent That $g_0^{(i)}$ is
site-independent can be proved  by considering that, for equal
charges and displacement amplitudes, the energies
$E^{(i)}=g_0^{(i)}n_i{\bf u}_i\bullet{\bf e}_{ij}$ $(i,j=1,2)$ on
both sites must coincide, i.e.
\beq
g_0^{(1)}{\bf u}_1\bullet{\bf
e}_{12}=g_0^{(2)}{\bf u}_2\bullet{\bf e}_{21} =-g_0^{(2)}{\bf
u}_2\bullet{\bf e}_{12} \label{g0equal}
\eeq
Now symmetry
requires $u_1=-u_2$ from which $g_0^{(2)}=g_0^{(1)}\equiv g_0$
follows.


\subsection{The crystal-field coupling}

\noindent This term expresses the change in the energy of
interaction between the charge on site $i$ and the potential
centred on the site $j$, that is:
\beq
\la\Psi_i^0|V_j(\R_j^0+u_j)|\Psi_i^0\ra -
\la\Psi^0_i|V_j(\R_{j}^0)|\Psi^0_i\ra\equiv g_{cf}^{(i)}n_iu_j
+\cal{O}(u_j^2)
\label{gcfnadef}
\eeq
In the literature, there is some confusion
on the form of this term, so we shall discuss it in some details,
following the line of reasoning used to prove that
$g_0^{(1)}=g_0^{(2)}$. Consider site $1$, with charge $n_{1}$.
Its energy, after a displacement ${\bf u}_{2}$, changes by an amount $%
E^{(1)}=g_{cf}^{(1)}n_{1}$  ${\bf u}_{2}\bullet {\bf e}_{12}$. \ This can
be considered as the quantity measured by an observer sitting on ion $1$ and
watching the ion $2$ \ moved by ${\bf u}_{2}$ .\ The equivalent measurement
done by  an observer \ on ion $2$ watching the ion $1$ displaced by ${\bf u}%
_{1}$,\ yields $E^{(2)}=g_{cf}^{(2)}n_{2}{\bf u}_{1}\bullet {\bf
e}_{21}$.
 Notice that each observer uses its own unit vector, pointing from the
observer's ion towards the other, displaced ion.  Formally,  $E^{(2)}$ can
be obtained by performing a site label permutation  on $E^{(1)}.$  An
external observer would find it more convenient to refer both quantities to
the same reference system, \ ${\bf e}_{12}$ say, yielding$\
E^{(2)}=-g_{cf}^{(2)}n_{2}$ ${\bf u}_{1}\bullet {\bf e}_{12}.$ Assuming
equal charge density  and displacement amplitude in the two observations,
one must have  $E^{(1)}=E^{(2)}.$ \ It follows, dropping the charge
densities:
\begin{equation}
g_{cf}^{(1)}{\bf u}_{2}\bullet {\bf e}_{12}=-g_{cf}^{(2)}{\bf u%
}_{1}\bullet {\bf e}_{12}\Longrightarrow
g_{cf}^{(1)}u_{1}=-g_{cf}^{(2)}u_{2}\Longrightarrow
g_{cf}^{(1)}=g_{cf}^{(2)}=g_{cf}^{{}}, \label{cfdef1}
\end{equation}
because the only displacements allowed by symmetry are such that \ $%
u_{1}=-u_{2}$. Therefore for the dimer as a whole one writes this term as
\begin{equation}
g_{cf}^{{}}\sum_{\sigma}(n_{1\sigma}u_{2}+n_{2\sigma}u_{1}).
\label{cfdef2}
\end{equation}
Passing now to the explicit evaluation for site $1$ we have:

\beq
g_{cf}^{(1)}=\lim_{u_2^{~}\rta 0} {{\partial } \over{
\partial u_2}}
\int^\infty_{-\infty}\Psi_1(\RIo,\RJo,x,y,z)^2V_2^{~}(u_2^{~})dxdydz.
\label{cfna1} \eeq
We change variables from $x$ to
$p=x/a/2-u_2^{~}$ and proceed in strict analogy to the evaluation
of $g_0^{(1)}$, arriving at:
\beq
g_{cf}^{(1)}=
2\int^\infty_{-\infty}\Psi_1{{\partial \Psi_1} \over{ \partial
x}}V_2 dxdydz. \label{cfna2}
\eeq
By explicitating $\Psi_1$ and
its derivative we obtain:
\[
g_{cf}^{(1)}=-4\Gamma^2\int^{\infty}_{-\infty}
\left(A^2\phi_1^2+B^2\phi_2^2 +2AB\phi_1\phi_2\right)xV_2 dxdydz
\]
\beq - 2\Gamma^2a\int^{\infty}_{-\infty}
\left(A^2\phi_1^2-B^2\phi_2^2\right)V_2dxdydz. \label{cfna3}
\eeq
>From the matrix elements evaluated in Appendix 2 we obtain:

\beq
g^{(1)}_{cf}=-2A\left({{\Gamma}\over{a}}\right)\sqrt{{{2}\over{\pi}}}
\left[AF_0(2a^2\Gamma^2)+4BSF_0(a^2\Gamma^2/2)-4BS^2-AS^4\right].
\label{g1cfnadexpl} \eeq

\vfill\eject
\subsection{The Su-Schrieffer-Heeger coupling}

\noindent This term is a non-diagonal inter-site coupling,
due to the modulation of
the hopping amplitude $t$.
As the Wannier functions keep being centred on the static lattice
positions, the difference in kinetic  energy caused by the displacements
is due only to the  contribution to $t$ from $V_1$ and $V_2$:
\[
\int \Psi_1(\rv-\RIo)[V_1(\rv-\RIo-\uI)+V_2(\rv-\RJo-\uJ)]\Psi_2(\rv-\RJo)
d^3\rv
\]
\[
 -\int \Psi_1(\rv-\RIo)[V_1(\rv-\RIo)+V_2(\rv-\RJo)]\Psi_2(\rv-\RJo)
d^3\rv
\]
\beq \equiv \gamma_{12}\sum_\sigma (c_{1\sigma}^\dagger
c^{~}_{2\sigma}+c_{2\sigma}^\dagger
c^{~}_{1\sigma})(u_2-u_1)+{\cal{O}}[(u_2-u_1)^2]. \label{g12na1}
\eeq Notice that, to preserve the invariance of the Hamiltonian
under site permutation, the $SSH$ coupling has to be odd under the
same operation: $\gamma_{12}=-\gamma_{21}$ (see e.g Refs. 2,3).\par
Let us distinguish
the two contributions as $t_{V_{1}}^{~}$ and $t_{V_{2}}^{~}$,
defined as:
\[
t_{V_{1}}^{~} =\int^{\infty}_{-\infty}
{{\Psi_1^{~}(u=0)\Psi_2^{~}(u=0)}\over{\sqrt{(x+a/2-u_1^{~})^2+y^2+z^2}}}
dxdydz,
\]
\begin{equation}
t_{V_{2}}^{~} =\int^{\infty}_{-\infty}
{{\Psi_1^{~}(u=0)\Psi_2^{~}(u=0)}\over{\sqrt{(x-a/2-u_2^{~})^2+y^2+z^2}}}
dxdydz.
\label{tv1v2}
\end{equation}
To evaluate $t_{V_1}^{~}$, let us define $p=x+a/2-u_1^{~}$ and use the
symmetry relation $u_1^{~}=-u/2$, so that we can derive the kernel:
\begin{equation}
{{\partial t_{V_1}^{~}}\over{\partial u}}=\int^{\infty}_{-\infty}
{{\partial [\Psi_1^{~}(p,u,y,z)\Psi_2^{~}(p,u,y,z)]}\over{\partial
u}}V_1^{~}(p,y,z)dpdydz.
\label{dtudu}
\end{equation}
Next we notice that
\begin{equation}
\lim_{u\rta 0} {{\partial
[\Psi_1^{~}(p,u,y,z)\Psi_2^{~}(p,u,y,z)]}\over{\partial u}}
=-{{1}\over{2}}\lim_{u\rta 0} {{\partial
[\Psi_1^{~}(p,u,y,z)\Psi_2^{~}(p,u,y,z)]}\over{\partial p}}
=-{{1}\over{2}}{{\partial (\Psi_1^{~}\Psi_2^{~})}\over{\partial
x}},
\label{psider}
\end{equation}
so that finally we can write:
\begin{equation}
\lim_{u\rta 0}{{\partial t^{~}_{V_1}}\over{\partial u}}=
-{{1}\over{2}}\int^{\infty}_{-\infty} {{\partial
[\Psi_1^{~}\Psi_2^{~}]}\over{\partial x}}V_1^{~}dxdydz.
\label{dtv1du}
\end{equation}
By performing similar manipulation on
$t_{V_2}^{~}(u_2)$, but now with $u_2^{~}=u/2$, we arrive at:
\begin{equation}
\lim_{u\rta 0}{{\partial t^{~}_{V_2}}\over{\partial u}}=
{{1}\over{2}}\int^{\infty}_{-\infty} {{\partial
[\Psi_1^{~}\Psi_2^{~}]}\over{\partial x}}V_2^{~}dxdydz.
\label{dtv2du}
\end{equation}
Summing the two contributions yields:
\beq
\gamma_{12}=-{1 \over
2}\int^\infty_{-\infty}dxdydz \left[{{\partial (\Psi_1\Psi_2)}
\over{ \partial x}} \right]V_1 +{1 \over
2}\int^\infty_{-\infty}dxdydz \left[{{\partial (\Psi_1\Psi_2)}
\over{ \partial x}}\right] V_2 \equiv -\cal{X}+\cal{Y}.
\label{g12na2}
\eeq
The first integral, $\cal{X}$, after
explicitating $\Psi_1$  and $\Psi_2$ becomes:
\beq
{\cal{X}}\equiv
{1 \over 2}\int^\infty_{-\infty}dxdydz {{\partial } \over {
\partial x}}\left[AB(\phi_1^2+\phi_2^2)+
(A^2+B^2)\phi_1\phi_2\right]V_1(x,y,z). \label{C1}
\eeq
The
derivatives:
\beq
{{\partial (\phi_1\phi_2)} \over { \partial
x}}=-4\Gamma^2x\phi_1\phi_2, \qquad {{\partial \phi_{1}^2} \over {
\partial x}}=-4\Gamma^2\left(x+{a \over 2}\right)\phi_{1}^2, \qquad
{{\partial \phi_{2}^2} \over { \partial x}}=-4\Gamma^2\left(x-{a
\over 2}\right)\phi_{2}^2. \label{C2} \eeq
when substituted into
\Eq{\ref{C1}} yield
\beq
{\cal{X}}= -2\Gamma^2
\left\{AB[\la\phi_1|(x+{a \over 2})V_1|\phi_1\ra + \la\phi_2|(x-{a
\over 2})V_1|\phi_2\ra] +(A^2+B^2)\la
\phi_1|xV_1|\phi_2\ra\right\}. \label{g12na3}
\eeq
 Similarly we get
\[
{\cal{Y}} \equiv
 {1 \over 2}\int^\infty_{-\infty}dxdydz
\left[{{\partial (\Psi_1\Psi_2)} \over{ \partial x}}\right]V_2,
\]
\beq
{\cal{Y}}=-2\Gamma^2\left\{ AB\left[\la\phi_1|(x+{a \over
2})V_2|\phi_1\ra +\la\phi_2|(x-{a \over 2})V_2|\phi_2\ra \right]
+(A^2+B^2)\la\phi_1|xV_2|\phi_2\ra\right\}. \label{C3}
\eeq
The
contributions from $\la\phi_1|(x+a/2)V_1|\phi_1\ra $ in
Eq.\ref{g12na3} and $\la\phi_2|(x-a/2)V_2|\phi_2\ra$ in
Eq.\ref{C3} vanish due to the
 parity of the kernel.
Substituting the matrix elements from Appendix 2 we get:
\[
\gamma_{12}=4\sqrt{{{2}\over{\pi}}}\left({{\Gamma}\over{a}}\right)
\left\{{{AB}\over{2}}\left[S_0^4-(1-4a^2\Gamma^2)F_0(2a^2\Gamma^2)\right]
\right\}
\]
\beq
+4\sqrt{{{2}\over{\pi}}}\left({{\Gamma}\over{a}}\right)
\left\{(A^2+B^2)S_0\left[S_0-F_0(a^2\Gamma^2/2)\right]\right\}.
\label{ga12nadexpl}
\eeq
Under site permutation $a\rta -a$ and
$A\rta B$ so that $\gamma_{12}=-\gamma_{21na}$ as expected.

In conclusion, in the non-adiabatic limit the electron-phonon
Hamiltonian is given by:

\[
H^{na}_{el-phon}= g_0\sum_\s (\nIs u_1+\nJs u_2)+
g_{cf}^{{}}\sum_{\sigma}(n_{1\sigma}u_{2}+n_{2\sigma}u_{1})
\]
\beq +\gamma_{12}\sum_\sigma (c_{1\sigma}^\dagger
c^{~}_{2\sigma}+c_{2\sigma}^\dagger c^{~}_{1\sigma})(u_2-u_1).
\eeq
\section{Couplings in the adiabatic limit}

One can use the method introduced in the non-adiabatic limit to
evaluate the couplings in the adiabatic case. However, one readily
verifies that, in the explicit expression of the different electronic
 interactions in the adiabatically displaced state, $u$
invariably enters in the combination $a+u$. Therefore the procedure of
first deriving the
integral kernels with respect to $u$, then taking the limit $u\rta 0$, and
finally evaluating the integrals is equivalent to first
 evaluating the interactions
for $u=0$, and then deriving them with respect to $a$. All the
parameters in Eq.\ref{helbare} were explicitly evaluated in Ref.1,
so we shall simply derive them with respect to the dimer length
$a$. \par

\subsection{The coupling term derived from $\epsilon$}

There is some confusion in the literature about the correct form
of the electron-phonon Hamiltonian obtained in this limit from the
variation of the local energy $\epsilon$, therefore we shall
devote some space to discussing this point. \par \noindent In the
adiabatic limit, $u_i\ne 0$ in  both the charge distributions and
in the potentials. As the origin of the $x$-coordinate can be
placed onto one of the displaced ions, this interaction couples
the charge on site $i$ to the position of site $j$ through the
modification of both the kinetic and the potential contributions.
One has then to take
 into account the {\it relative} displacement of the ions.
To obtain coupling terms from  $\epsilon_{\nabla}$ and
$\epsilon^{(i)}_{V_{i}}$ it is essential that one uses the proper
Wannier functions. If one instead adopted the local orbitals, such
couplings would vanish. In the adiabatic limit, therefore, the
overall $\epsilon$-derived electron-phonon coupling term in the
Hamiltonian is:
\beq
g_{\epsilon}^{(1)}\sum_\s n_{1\s}(u_2-u_1)
+g_{\epsilon}^{(2)}\sum_\s n_{2\s}(u_1-u_2).
\label{gcfaddef}
\eeq
\noindent As $g_{\epsilon}^{(1)}=g_{\epsilon}^{(2)}$
 we can drop the site indexes, and write the total adiabatic
contribution from local energy terms to the electron-phonon
Hamiltonian as:
\beq
H^{\epsilon}_{el-phon}=
g_{\epsilon}\sum_\s(\nJs-\nIs)(u_1-u_2).
 \label{cfad13}
\eeq An expression similar to Eq.\ref{cfad13} has been proposed in
Ref.7. To convince the skeptical reader, in Appendix 4 we show,
using
 the contribution from $\epsilon_{V_{2}}^{(1)}$ as an
example, that, starting from the general definition of the coupling,
one  verifies Eq.\ref{cfad13}.\par

\subsection{Explicit expressions for the adiabatic couplings.}

\subsubsection{Couplings originated from one-body electronic
interactions.}

There are two type of couplings: one ($g_{\epsilon}$),
 derived from the local energy:
\begin{equation}
g^{~}_{\epsilon}\equiv g_{\nabla}^{(i)}+ g_{V_{i}}^{(i)}+
g_{V_{j}}^{(i)} \qquad (i,j=1,2),
 \label{epgeps}
\end{equation}
and one ($\gamma_{12}$) from the hopping term
(Su-Schrieffer-Heeger coupling):
\beq
\gamma_{12}\equiv
\gamma_{12\nabla}+\gamma_{12V}.
 \label{gam12def}
\eeq
Below we list each  contribution to the electronic
interactions and the derived electron-phonon couplings. From the
local energy on site $1$
\[
\epsilon^{(1)} \equiv \epsilon_{\nabla}^{~} +\epsilon_{V_1}^{(1)}
+\epsilon_{V_2}^{(1)},
\]
\[
\epsilon^{~}_{\nabla} ={{\hbar^2}\over{2m}}\left[3\Gamma^2+
\Gamma^4\left({{a^2S^2}\over{1-S^2)}}\right)\right],
\]

\[
\epsilon_{V_{1}}^{(1)}=-Ze^2\left(2\Gamma\sqrt{ {{2}\over{\pi}}
}\right)
\left[A^2+B^2F_0(2a^2\Gamma^2)+2ABSF_0(a^2\Gamma^2/2)\right],
\]
\beq
\epsilon_{V_2}^{(1)}=-Ze^2\left(2\Gamma\sqrt{{{2}\over{\pi}}}\right)
\left[B^2+A^2F_0(2a^2\Gamma^2)+2ABSF_0(a^2\Gamma^2/2)\right],
\label{eps3}
\eeq
one obtains the three contributions to
\[
g_{\epsilon}^{(1)}\equiv
g_{\nabla}^{~}+g_{V_1}^{(1)}+g_{V_2}^{(1)},
\]
\[
 g_{\nabla}=
-{{\hbar^2}\over{2m}}\left[{{a\Gamma^4S^2}\over{(1-S^2)^2}}\right]
\left[2\left(1-a^2\Gamma^2-S^2\right)\right],
\]
\[
 g^{(1)}_{V_1}=
-Ze^2\left(2\Gamma\sqrt{{{2}\over{\pi}}}\right) \left[{{\partial
A^2}\over{\partial u}}+{{2ABS^2+B^2S^4}\over{a}}
+F_0(2a^2\Gamma^2)\left({{\partial B^2}\over{\partial u}}-
{{B^2}\over{a}}\right)\right],
\]
\[
-Ze^2\left(2\Gamma\sqrt{{{2}\over{\pi}}}\right)
(2S)\left[{{\partial AB}\over{\partial
u}}-{{AB}\over{a}}\left(1+a^2\Gamma^2)\right)\right],
\]
\[
 g^{(1)}_{V_2}=-Ze^2\left(2\Gamma\sqrt{{{2}\over{\pi}}}\right)
\left[{{\partial B^2}\over{\partial u}}+{{2ABS^2+A^2S^4}\over{a}}
+F_0(2a^2\Gamma^2)\left({{\partial A^2}\over{\partial u}}-
{{A^2}\over{a}}\right)\right],
\]
\beq
-Ze^2\left(2\Gamma\sqrt{{{2}\over{\pi}}}\right)
(2S)\left[{{\partial AB}\over{\partial u}}
-{{AB}\over{a}}\left(1+a^2\Gamma^2)\right)\right].
 \label{geps}
\eeq
The two potential contributions depend on the site for which
they are evaluated, but their sum,  $g^{(1)}_{V_1}+g^{(1)}_{V_2}$,
does not.\par For the hopping term,  we can lump together the
potential contributions which individually have no particular
physical meaning. Then from
\[
t\equiv t_{\nabla}^{~}+t_V^{~},
\]
\[
t_{\nabla}={{\hbar^2}\over{2m}}{{a^2S\Gamma^4}\over{(1-S^2)}},
\]
\beq t_{V}= -Ze^2\left(2\Gamma\sqrt{{{2}\over{\pi}}}\right)
\left\{2AB\left[1+F_0(2a^2\Gamma^2)\right]+2(A^2+B^2)
SF_0(a^2\Gamma^2/2)\right\},
\label{tnablaV}
\eeq
the
Su-Schrieffer-Heeger coupling $\gamma^{~}_{12}\equiv
\gamma^{~}_{12\nabla}+\gamma^{~}_{12V}$ is obtained as:
\[
\gamma_{12\nabla}=
{{\hbar^2}\over{2m}}\left[{{aS\Gamma^4}\over{(1-S^2)^2}}\right]
\left[2(1-S^2)-a^2\Gamma^2(1+S^2)\right],
\]
\[
\gamma_{12V}=
-Ze^2\left(4\Gamma\sqrt{{{2}\over{\pi}}}\right)\Bigg\{
\left[{{\partial AB}\over{\partial u}}
+{{A^2+B^2}\over{a}}S^2+{{AB}\over{a}}S^4\right] +
\left[{{\partial AB}\over{\partial u}}
-{{AB}\over{a}}\right]F_0(2a^2\Gamma^2),
\]
\begin{equation}
+S\left[{{\partial (A^2+B^2)}\over{\partial u}}
-{{A^2+B^2}\over{a}}(1+a^2\Gamma^2)\right]F_0(a^2\Gamma^2/2)\Bigg\}.
\label{SSHV}
\end{equation}

Notice that, as the partial derivatives are linear in $a$, then both
$\gamma_{12\nabla}^{~}$ and $\gamma^{~}_{12V}$ change
 sign under site permutation, as expected from Refs.2,3.
\par
The electron-phonon Hamiltonian in the adiabatic limit has
therefore the form:

\beq
H^{ad}_{el-phon}=
g_{\epsilon}^{{}}\sum_{\sigma}(n_{1\sigma}-n_{2\sigma})(u_{2}-u_{1})
 +\gamma_{12}\sum_\sigma (c_{1\sigma}^\dagger
c^{~}_{2\sigma}+c_{2\sigma}^\dagger c^{~}_{1\sigma})(u_2-u_1).
\eeq

\subsubsection{Couplings originated from two-body electronic
interactions.}

Also the two-body electronic interactions $U,V,J(=P),X$ of Eq. (1)
 give origin to electron-phonon couplings which, to the best of our
knowledge, have never been considered in the literature up to now.
The electron-phonon coupling originating from the modulation of the
kinetic exchange in the $t-J$ Hamiltonian has been considered in
Ref.17. However, the kinetic exchange
$J_{kin}\equiv 2t^2/(U-V)$ is a composite
quantity different from the direct exchange $J_{xy}=J_z=P$ in Eq. (1).
Here we list the values the two-body electron-phonon interactions
 have in our model. They were obtained
by deriving with respect to $a$ the two-body interactions
evaluated explicitly in Ref.1. Notice that, as
$U(a)=J(a)+e^2\Gamma/\sqrt{\pi}$, then $dU/da=dJ/da=dP/da$
\[
{{dX}\over{da}}=-e^2{{\Gamma}\over{\sqrt{\pi}}}
\left[\left(-a\Gamma S\right)
 \frac{\left( 1+3S^{2}\right) }{(1-S^{2})^{3}}\right]
\left[1+2S^{2}+F_{0}\left( a^{2}\Gamma ^{2}\right)
 -2(1+S^{2})F_{0}\left( \frac{a^{2}\Gamma ^{2}}{4}\right) \right]
\]
\[
-e^{2}\frac{\Gamma }{\sqrt{\pi }}\left[ \frac{S/a}{(1-S^{2})^{2}}\right]
\left\{ 4a^{2}\Gamma ^{2}S^{2}\left[ F_{0}\left( \frac{a^{2}\Gamma ^{2}}{4}%
\right) -1\right] +S^{2}-F_{0}\left( a^{2}\Gamma ^{2}\right) \right\}
\]
\begin{equation}
-e^{2}\frac{\Gamma }{\sqrt{\pi }}\left[
\frac{S/a}{(1-S^{2})^{2}}\right] \left\{ 2(1+S^{2}) \left[
F_{0}\left( \frac{a^{2}\Gamma ^{2}}{4}\right) -\sqrt{S}\right]
\right\},   \label{dXda}
\end{equation}
\[
\frac{dU}{da}=
e^{2}\frac{\Gamma }{\sqrt{\pi }}\left[ \frac{-4a\Gamma
^{2}S^{2}}{(1-S^{2})^{3}}\right] \left[ 2-S^{2}+2S^{4}+S^{2}F_{0}\left(
a^{2}\Gamma ^{2}\right) -4S^{2}F_{0}\left( \frac{a^{2}\Gamma ^{2}}{4}\right)
\right]
\]
\[
+e^{2}\frac{\Gamma }{\sqrt{\pi }}\left[ \frac{S^{2}/a}{(1-S^{2})^{2}}\right] %
\Bigg\{2a^{2}\Gamma ^{2}\left[ 1-4S^{2}-F_{0}\left( a^{2}\Gamma ^{2}\right)
+4F_{0}\left( \frac{a^{2}\Gamma ^{2}}{4}\right) \right]
\]
\begin{equation}
+S^{2}-F_{0}\left( a^{2}\Gamma ^{2}\right) +4\left[ F_{0}\left( \frac{%
a^{2}\Gamma ^{2}}{4}\right) -\sqrt{S}\right] \Bigg\},
\label{dUda}
\end{equation}
\[
\frac{dV}{da}=e^{2}\frac{\Gamma }{\sqrt{\pi }}\left[- \frac{4a\Gamma
^{2}S^{2}}{(1-S^{2})^{3}}\right] \left[ 3-S^2-8S^4-(7-5S^2)F_0(a^2\Gamma^2)
-4(1-3S^2)F_0\left({{a^2\Gamma^2}\over{4}}\right)\right]
\]
\[
+e^{2}\frac{\Gamma }{\sqrt{\pi }}\left[ \frac{1/a}{(1-S^{2})^{2}}\right] %
\Bigg\{2a^{2}\Gamma ^{2}S^{2}\left[ -1-4S^{2}+F_{0}\left( a^{2}\Gamma
^{2}\right) +4F_{0}\left( \frac{a^{2}\Gamma ^{2}}{4}\right) \right]
\]
\begin{equation}
+2S^2+{{3}\over{2}}S^4-\left(2-{{3}\over{2}}S^2\right)F_0(a^2\Gamma^2)
+2S^{2} \left[ F_{0}\left( \frac{a^{2}\Gamma
^{2}}{4}\right)-\sqrt{S} \right] \Bigg\}, \label{dVda}
\end{equation}
\[
\frac{dJ}{da}\equiv\frac{dP}{da}\equiv\frac{dU}{da}. \label{dJda}
\]
All the above interactions change sign under site permutation. The
terms they contribute to the electron-phonon Hamiltonian all have
the same form, namely: \beq H_{Y}^{ep}= \frac
{dY}{da}F(c^{\dagger}_{i\sigma},c^{~}_{j\sigma}) (u_j-u_j)\qquad
(i,j=1,2) \eeq \noindent where $Y=U,V,X,J$ and
$F(c^{\dagger}_{i\sigma},c^{~}_{j\sigma})$ $(i,j=1,2)$ is the
function of Fermi operators representing the two-body interaction
whose amplitude is $Y$


\section{Results}

\noindent By definition,
 as $n_{i\s}$ is the number of electrons per site
(a dimensionless quantity), then the interaction parameters have
dimensions $[energy][length]^{-1}$. To get them in energy units
(eV) we shall measure  the deformations $u_i$  in units of the
characteristic phonon length $L=\sqrt{\hbar/(2\Omega M)}$. We
choose the phonon frequency such that $\hbar\Omega=0.1$ eV, which
is appropriate to $HTS$ and $CMR$, and $M$ equal to the mass of
$^{16}O$.
\begin{figure}
\centerline{\psfig{figure=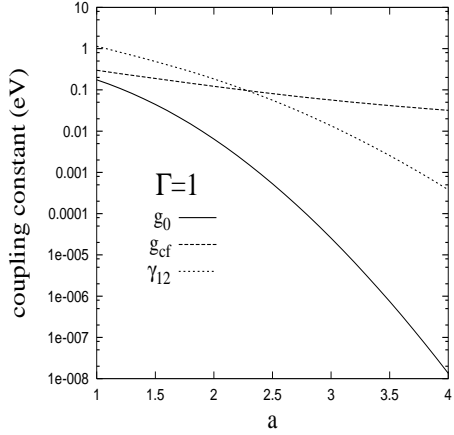,height=6cm,width=6cm}}
\caption {Non-adiabatic coupling constants
$g_0,g_{cf},\gamma_{12}$ (in eV) versus the dimer length (in \AA),
evaluated assuming $\Gamma=1.0 \AA^{-1}$.} \label{fig1}
\end{figure}
Figs.1 and 2 show the behaviour versus the dimer length $a$ (in \AA)
 of the non-adiabatic coupling parameters
$g_0,g_{cf},\gamma_{12}$ (in eV) evaluated by assuming two
representative values $\Gamma=1.0,2.0$\AA$^{-1}$  for the
shape-controlling parameter  of the Wannier functions. The  most
unexpected result concerns $g_{cf}$. While usually neglected in
the literature\cite{barisic} on metallic systems, this coupling
has been recognized as relevant to polar materials\cite{barisic2}.
We find indeed that, when $\Gamma=1.0$\AA$^{-1}$, $g_{cf}$ is
larger than $g_0$ for any $a$,
 and it becomes the largest
parameter for $a>2.2$\AA. For small $a$, the $SSH$ coupling is the
largest.
The strength of all the couplings decreases with $a$,
less quickly for $g_{cf}$, which, for large $a$, is still
comparable to the values of the hopping amplitude $t(a)$. \par
\begin{figure}
\centerline{\psfig{figure=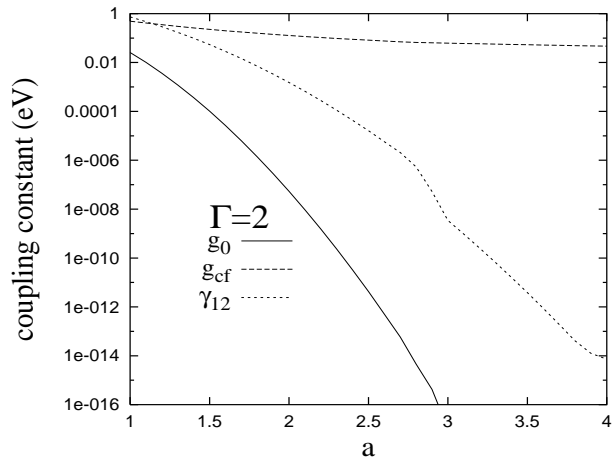,height=6cm,width=8cm}}
\caption {Non-adiabatic coupling constants $g_{cfad},
\gamma_{12ad}$ (in eV) versus the dimer length (in \AA), for
$\Gamma=2.0$\AA$^{-1}$.} \label{fig2}
\end{figure}
When $\Gamma=2.0$\AA$^{-1}$(see Fig.2) $g_0$ is negligible for any
$a$, $\gamma_{12}$ is large for small $a$ but drops very rapidly to
negligible values as $a$ increases, while $g_{cf}$ keeps appreciable
values for all $a$ values. We can conclude that, in the non-adiabatic
limit, the more localized
are the orbitals, the more relevant is the role of $g_{cf}$ in
relation to the other admissible couplings.\par
\begin{figure}
\centerline{\psfig{figure=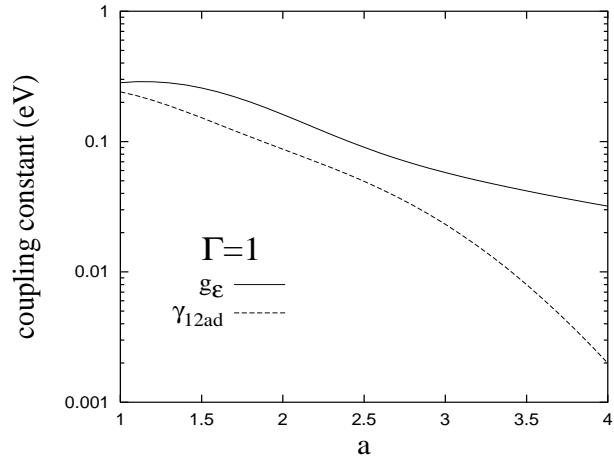,height=6cm,width=8cm}}
\caption {Adiabatic coupling constants $g_{\epsilon},
\gamma_{12ad}$ (in eV) versus the dimer length (in \AA), for
$\Gamma=1.0$ \AA$^{-1}$.} \label{fig3}
\end{figure}
Fig.3 is the adiabatic counterpart of Fig.1. The Holstein type
interaction is absent, being identically vanishing as discussed above.
 Here we find that $g_{\epsilon}$ is always larger than the $SSH$
interaction $\gamma_{12}$, and particularly for large $a$ there is
an order of magnitude difference between them. \par
\begin{figure}
\centerline{\psfig{figure=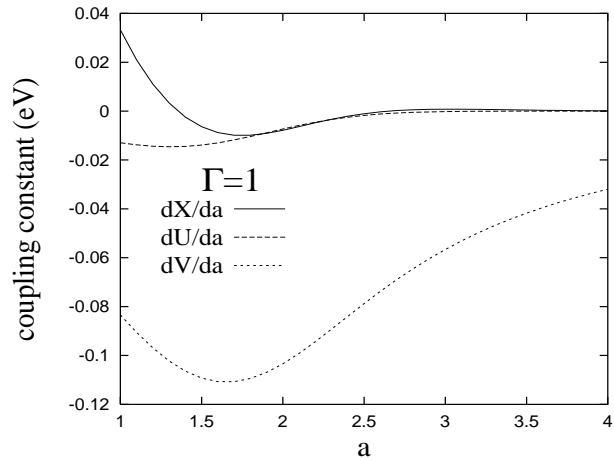,height=6cm,width=8cm}}
\caption {Adiabatic coupling constants from two-body interactions
 $dX/da, dU/da, dV/da$ (in eV) versus
the dimer length (in \AA), for $\Gamma=1.0$ \AA$^{-1}$.}
\label{fig4}
\end{figure}
Fig.4 shows the couplings derived from the two-body electronic
interactions for the same parameters as Fig.3. In general, their
values are smaller than those of $g_{\epsilon}$ and $\gamma_{12}$,
with the possible exception of $dV/da$. Indeed, that coupling
arises from a physical mechanism not very different from the one
originating $g^{(i)}_{V_j}$, i.e. the vibration of the charge on
site $j$ as felt by site $i$. Similarly to $g_{\epsilon}$ also
$dV/da$ decreases slowly with $a$, so that for large $a$ those two are
the only relevant couplings. It is worth stressing that, though
$U$ is larger than $J$, their derivatives coincide. Besides, the
derivative of the interaction coupling $X$ is a non-monotonic
function of the lattice constant. Indeed, $X$ exhibits a maximum
for $a\simeq$ 1.8 and then sharply decreases to a negligible
value.

\par
\begin{figure}
\centerline{\psfig{figure=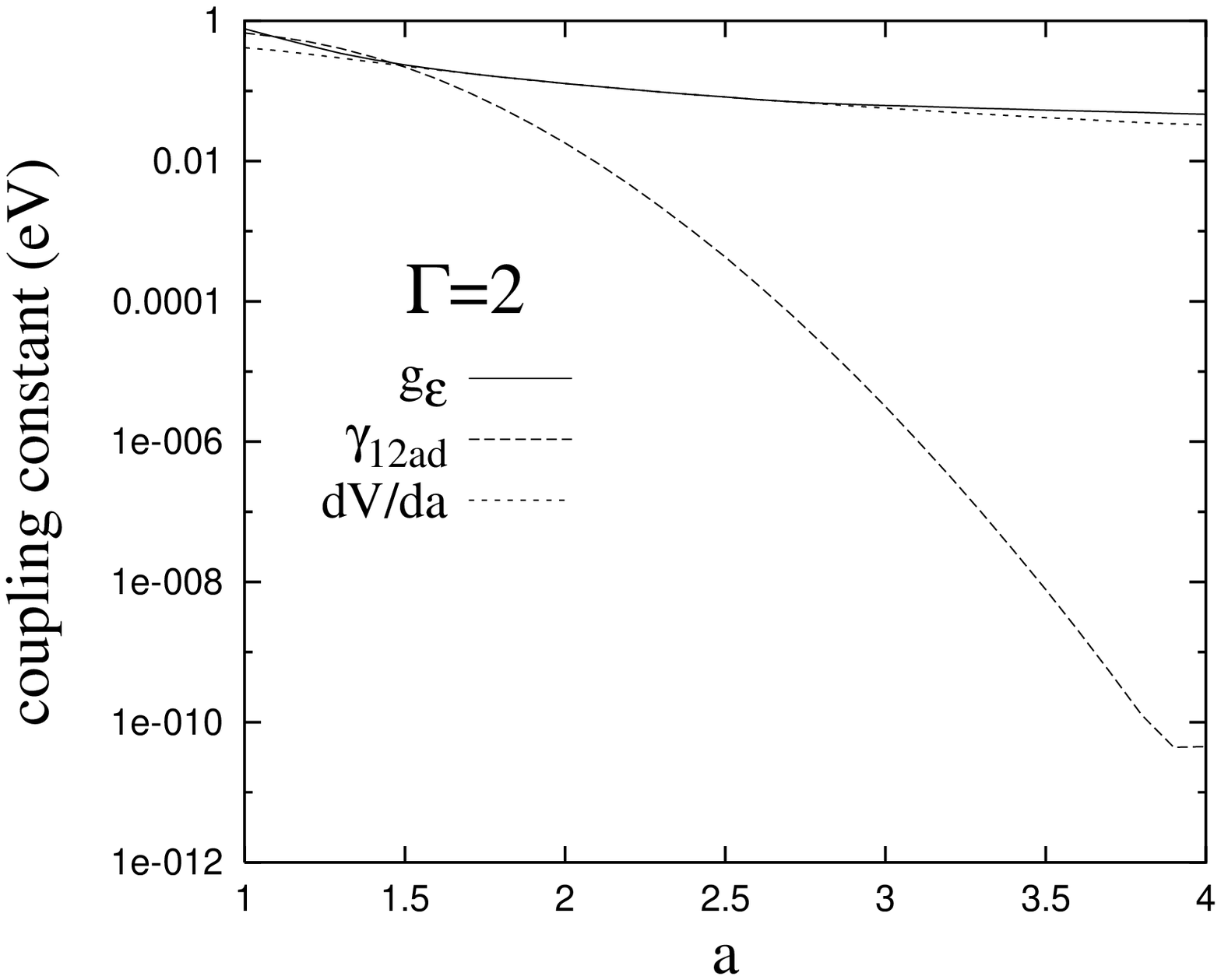,height=6cm,width=8cm}}
\caption {Adiabatic coupling constants $g_{\epsilon}, \gamma_{12},
dV/da$ (in eV) versus the dimer length (in \AA), for $\Gamma=2.0$
\AA$^{-1}$.} \label{fig5}
\end{figure}
When $\Gamma=2.0$\AA$^{-1}$, as shown in Fig.5, $g_{\epsilon}$ and
$\gamma_{12}$ have trends similar to the non-adiabatic case, with
larger values for small $a$. As $a$ increases, $\gamma_{12}$ quickly
reduces to negligible values, while $g_{\epsilon}$ decreases more slowly.
Of the two-body couplings, only $dV/da$ is non-negligible, and it has
a strength very close (in absolute value) to $g_{\epsilon}$.\par

\section{Conclusions}

\noindent We have presented the analytical evaluation of the electron-phonon
coupling parameters derived from both one- and two-body electronic interactions
 in a model of a dimer with a non-degenerate orbital
built from atomic orbital of Gaussian shape. The approach we have
followed (proposed in general terms in Refs.2,3)
 consists in inserting the site displacements $u_1,u_2$
 in the kernels of the
integrals defining the one-body electronic parameters $\epsilon_0$ and
$t$, and then considering the first-order terms in the expansion of
$\epsilon_0(u_1,u_2)$ and $t(u_1,u_2)$. In the adiabatic case
this procedure is equivalent to the simpler
one\cite{labbe,barisic,kuzemsky}$^{-}$\cite{petru}
 consisting in evaluating the
electronic parameters as functions of the lattice parameter
(corresponding to the dimer length $a$) and then deriving with
respect to $a$. In the anti-adiabatic case, however, the latter
procedure can not be applied, because $a$ and $u$ enter
independently the kernel of the integrals. From the physical point
of view, the "$a$-derivative" method describes how the electronic
parameters vary under a quasi-static (low frequency) change of the
equilibrium position of the ions, as could be realized e.g. under
pressure. In general the electron-phonon couplings are the
 effect of the ion oscillations at frequency $\Omega$, which causes the
system to be alternatively compressed and elongated over a time
$\approx \Omega^{-1}$. It is this implicit dynamics of the
displacements which allows for distinguishing between the
adiabatic and non-adiabatic regimes, and forbids the use of the
"$a$-derivative" method in the latter case.\par A novel result is
the evaluation of the couplings originating from the two-body
electronic interactions. We have shown that at least one of them,
generated by the Coulomb repulsion between the charges on
different sites, is comparable, or even larger, than the couplings
derived from the one-body interactions.\par The quantitative
results for  the coupling parameters, even if agreeing in order of
magnitude with some estimates from experimental data (see, e.g.
Ref.13) are obviously model-depending. However, their ratios
should be more close to the reality. In particular, the obtained
values of the various couplings, when compared to the values of
the electronic interactions obtained from the same Wannier
functions\cite{acquarone}
 suggests that, for dimer lengths comparable to the lattice parameters
in $HTS$ and $CMR$, only $dU/da$ and $dX/da$ can be safely dropped,
while neglecting {\it any} of the other
electron-phonon interactions is a questionable approximation. \par
Finally, we have been able to determine the correct form of the
admissible electron-phonon coupling one-body terms in the
interacting Hamiltonian. In the literature one sometimes finds
proposals for such terms which are incompatible with our results.
This is the case, for instance, of the electron-phonon Hamiltonian
of Refs.14,15.

\nonumsection{Acknowledgements}
\noindent It is a pleasure to
thank J.R. Iglesias, M. A. Gusm\~{a}o,
 A. Painelli, M. Cococcioni, A. Alexandrov, and particularly A.A. Aligia,
for critical discussions and comments. This work was supported by
I.N.F.M. and by MURST 1997 co-funded project "Magnetic Polarons in
Manganites" .
\par
\vfill\eject
\section{Appendix 1}

\noindent Here we list some results of frequent use in the
calculations. We shall use the zero-displacement derivatives of
$A$ and $B$: \beq {{\partial A} \over{ \partial u}}\Bigg|_{u \rta
0}= {{a \Gamma^2 S_0}\over{4}} \left[ - (1+S_0)^{-3/2}+
(1-S_0)^{-3/2} \right], \label{wandadu}
\eeq
\beq {{\partial B}
\over{ \partial u}}\Bigg|_{u \rta 0}= {{a \Gamma^2 S_0}\over{4}}
\left[ - (1+S_0)^{-3/2}- (1-S_0)^{-3/2}\right].
\label{wandbdu}
\eeq
\noindent In the following we shall use the substitution
$t=x+a/2-u_1$ in both $\Psi_1$ and $\Psi_2$, yielding:
\[
\Psi_1(x, u_1,u_2,y,z)\Rightarrow \Psi_1(t, u,y,z)=
 A(u)Ne^{-\Gamma^2[t^2+y^2+z^2] }
+ B(u)Ne^{-\Gamma^2[(t-a+u)^2+y^2+z^2]},
\]
\beq \Psi_2(x,u_1,u_2,y,z)\Rightarrow \Psi_2(t, u,y,z)=
B(u)Ne^{-\Gamma^2[t^2+y^2+z^2]} +
A(u)Ne^{-\Gamma^2[(t-a+u)^2+y^2+z^2]}. \label{wanpsi} \eeq To
write down their derivatives in the zero-displacement limit, it is
convenient to revert to $x=t-a/2$ so that: \beq {{\partial
\Psi_1(x,y,z,u)} \over{\partial u}}\Bigg\vert_{u\rta 0}
=\phi_1(x,y,z) {{\partial A} \over{\partial u}} + \phi_2(x,y,z)
\left[ {{\partial B} \over{\partial u}} -2\Gamma^2(x-{a\over
2})B\right], \label{dpsi1dub} \eeq \beq {{\partial
\Psi_2(x,y,z,u)} \over{\partial u}}\Bigg\vert_{u\rta 0}
=\phi_1(x,y,z) {{\partial B} \over{\partial u}} + \phi_2(x,y,z)
\left[ {{\partial A} \over{\partial u}} -2\Gamma^2(x-{a\over
2})A\right]. \label{dpsi2dub} \eeq We shall also use the other
substitution $t\equiv x-a/2-u_2$, yielding:
\[
\Psi_1(x, u_1,u_2,,y,z)\Rightarrow
A(u)Ne^{-\Gamma^2[(t+a-u)^2+y^2+z^2] } +
B(u)Ne^{-\Gamma^2[t^2+y^2+z^2]},
\]
\beq \Psi_2(x,u_1,u_2,y,z)\Rightarrow
B(u)Ne^{-\Gamma^2[(t+a-u)^2+y^2+z^2]} +
A(u)Ne^{-\Gamma^2[t^2+y^2+z^2]}. \label{wanpsi2}
\eeq
In the limit
of vanishing deformations  it follows, after reverting
 to $x=t+a/2$:
\beq {{\partial \Psi_1(x,y,z,u)} \over{\partial
u}}\Bigg\vert_{u\rta 0} = \phi_1(x,y,z) \left[ {{\partial A}
\over{\partial u}} +2\Gamma^2(x+{a\over 2})A\right] +\phi_2(x,y,z)
{{\partial B} \over{\partial u}}, \label{dpsi1dub2} \eeq \beq
{{\partial \Psi_2(x,y,z,u)} \over{\partial u}}\Bigg\vert_{u\rta 0}
= \phi_1(x,y,z) \left[ {{\partial B} \over{\partial u}}
+2\Gamma^2(x+{a\over 2})B\right] +\phi_2(x,y,z) {{\partial A}
\over{\partial u}}. \label{dpsi2dub2} \eeq
\section{Appendix 2}

\noindent For convenience reasons we shall  list here various
matrix elements, all evaluated  in the limit of vanishing
deformations, which enter the calculations. Those of them which do
not follow straightforwardly from standard properties of Gaussian
integrals\cite{rizhik}
  are evaluated in details in the Appendix 3. By defining:
\beq F_0(x)={{1} \over{ \sqrt{x} }}\int^{\sqrt{x}}_0e^{-t^2}dt,
\label{hol27} \eeq we have: \beq \la
\phi_i|-\nabla^2/2|\phi_i\ra={{3\Gamma^2} \over {2}}+
{{\Gamma^4a^2S_0^2} \over{2(1-S_0^2)}}, \qquad \la
\phi_i|-\nabla^2/2|\phi_j\ra=-{{\Gamma^4a^2S_0}
\over{2(1-S_0^2)}}, \qquad (i,j=1,2) \label{matel1} \eeq \beq \la
\phi_i|V_i|\phi_i\ra=2\Gamma\sqrt{2/\pi}, \qquad \la
\phi_i|V_j|\phi_i\ra=2\Gamma\sqrt{2/\pi}F_0(2a^2\Gamma^2),
\label{matel1b} \eeq \beq \la
\phi_i|V_i|\phi_j\ra=2\Gamma\sqrt{2/\pi}S_0F_0
\left({{a^2\Gamma^2}\over{2}}\right), \label{matel2} \eeq \beq \la
\phi_1|xV_1|\phi_1\ra=-a\Gamma\sqrt{2/\pi}, \qquad \la
\phi_2|xV_2|\phi_2\ra=a\Gamma\sqrt{2/\pi}, \label{matel3} \eeq
\[
\la\phi_1|xV_2|\phi_1\ra= {{1}\over{2a\Gamma}}\sqrt{{2\over \pi} }
\left[-S_0^4+ \left(1-2a^2\Gamma^2 \right)F_0(2a^2\Gamma^2)
\right],
\]
\beq \la\phi_2|xV_1|\phi_2\ra=-\la\phi_1|xV_2|\phi_1\ra,
\label{matel4} \eeq
\[
\la\phi_1|xV_1|\phi_2\ra= {{ S_0\sqrt{2/\pi} } \over{a\Gamma
}}\left[S_0 -F_0\left({{a^2\Gamma^2}\over{2}}\right)\right],
\]
\beq \la\phi_1|xV_2|\phi_2\ra=-\la\phi_1|xV_1|\phi_2\ra.
\label{matel5} \eeq Some other useful relations are: \beq \la
\phi_1|V_1|\phi_1\ra = \la \phi_2|V_2|\phi_2\ra, \qquad \la
\phi_2|V_1|\phi_2\ra = \la \phi_1|V_2|\phi_1\ra. \label{hol12}
\eeq We can directly evaluate \beq \la\phi_1|xV_1|\phi_1\ra = -a
\Gamma \sqrt{2/\pi}= - \la\phi_2|xV_2|\phi_2\ra. \label{hol13}
\eeq Indeed, due to the odd parity of the kernel,
$\la\phi_1|(x+a/2)V_1|\phi_1\ra =0$ from which \Eq{\ref{hol13}}
follows. We can also prove that \beq \la\phi_1|xV_2|\phi_2\ra = -
\la\phi_2|xV_1|\phi_1\ra, \qquad \la\phi_1|xV_2|\phi_1\ra = -
\la\phi_2|xV_1|\phi_2\ra. \label{hol14} \eeq Indeed, if $x$
changes sign, then $V_1\rta V_2$ and $\phi_1\rta\phi_2$. Therefore
the functions $\phi_1(V_1+V_2)\phi_2$ and $  \phi_1^2V_2+
\phi_2^2V_1$ are  even in $x$, so that:
\[
\la\phi_1|xV_2|\phi_2\ra + \la\phi_2|xV_1|\phi_1\ra =
\la\phi_1|x(V_1+V_2)|\phi_2\ra= 0,
\]\beq
\la\phi_1|xV_2|\phi_1\ra + \la\phi_2|xV_1|\phi_2\ra =
\int^\infty_{-\infty} x(\phi_1^2V_2+ \phi_2^2V_1)dxdydz =0,
\label{hol15} \eeq because the integrands are odd functions of x.

\vfill\eject
\section{Appendix 3}

\noindent Let us now evaluate the unknown integrals in \Eq{\ref{hol6}}. To do
that, consider
\[
\la\Psi_1|V_1|\Psi_1\ra=
\int^\infty_{-\infty}\Psi_1(a/2,x,y,z)^2V_1(a/2,x,y,z)dxdydz,
\]
and its derivative with respect to $a/2$:
\[
{{\partial\la\Psi_1|V_1|\Psi_1\ra} \over{\partial(a/2)}}=
2\int^\infty_{-\infty}\Psi_1
\left[ {{\partial \Psi_1} \over{ \partial(a/2)}} \right]
V_1dxdydz
+\int^\infty_{-\infty}\Psi_1^2
{{\partial V_1} \over{ \partial(a/2)}} dxdydz
\]
\beq = 2\int^\infty_{-\infty}\Psi_1 \left[ {{\partial \Psi_1}
\over{ \partial(a/2)}} \right] V_1dxdydz - g_0^{(1)},
\label{hol17} \eeq where we use the \Eq{\ref{hol4}} and
$\lim_{u_1\rta 0} {{\partial V_1} / {\partial u_1}}=
 -{{\partial V_1} / {\partial x}}= -{{\partial V_1} / {\partial (a/2)}}$.
 To evaluate the first integral we need also:
\beq {{\partial \Psi_1} \over{ \partial(a/2)}} ={{\partial A}
\over{ \partial(a/2)}}\phi_1 + {{\partial B} \over{
\partial(a/2)}} \phi_2 - 2
\Gamma^2\left[A\left(x+a/2\right)\phi_1-B(x-a/2)\phi_2\right].
\label{hol18} \eeq Inside the integral, it is convenient to use
$\partial/\partial x$,
 instead of $\partial/\partial(a/2)$.
As \beq {{\partial \Psi_1} \over{ \partial x}}= -2 \Gamma^2
\left[A\left(x+a/2\right)\phi_1+B(x-a/2)\phi_2\right],
\label{hol19} \eeq then \beq {{\partial \Psi_1} \over{
\partial(a/2)}} ={{\partial A} \over{ \partial(a/2)}}\phi_1 +
{{\partial B} \over{ \partial(a/2)}} \phi_2 +{{\partial \Psi_1}
\over{ \partial x}} +4\Gamma^2B(x-a/2)\phi_2, \label{hol20} \eeq
so that \Eq{\ref{hol17}} becomes, recalling \Eq{\ref{hol4}}:
\[
{{d\la\Psi_1|V_1|\Psi_1\ra} \over{d(a/2)}}=
 2{{\partial A} \over{ \partial(a/2)}}
\int^\infty_{-\infty}\Psi_1\phi_1V_1 dxdydz
+ 2{{\partial B} \over{ \partial(a/2)}}
\int^\infty_{-\infty}\Psi_1\phi_2V_1 dxdydz
\]
\beq +8\Gamma^2B\int^\infty_{-\infty}\Psi_1(x-a/2)\phi_2V_1dxdydz.
\label{hol21} \eeq Developing $\Psi_1=A\phi_1+B\phi_2$
 and reordering, one arrives at:
\[
8\Gamma^2 \left(
AB\la\phi_1|xV_1|\phi_2\ra+B^2\la\phi_2|xV_1|\phi_2\ra \right)=
{{\partial \la\Psi_1|V_1|\Psi_1\ra} \over{\partial (a/2)}}
-{{\partial A^2} \over{ \partial(a/2)}}\la\phi_1|V_1|\phi_1\ra
\]
\beq +\left[4a\Gamma^2B^2- {{\partial B^2} \over{ \partial(a/2)}}
\right]\la\phi_2|V_1|\phi_2\ra + \left[4a\Gamma^2AB -2{{\partial
(AB)} \over{ \partial(a/2)}} \right]\la\phi_1|V_1|\phi_2\ra.
\label{hol22} \eeq This equation connects the unknown integrals on
the left hand side to known ones. We need another relation, which
is provided by similar manipulations on $\la\Psi_2|V_1|\Psi_2\ra$.
\[
{{ \partial\la\Psi_2|V_1|\Psi_2\ra} \over{ \partial(a/2)}}
={{ \partial} \over{ \partial(a/2)}}\left [\int^\infty_{-\infty}
\Psi_2(a/2,x,y,z)^2V_1(a/2,x,y,z)dxdydz \right]
\]
\beq
=2\int^\infty_{-\infty} \Psi_2 {{ \partial\Psi_2} \over{
\partial(a/2)}}V_1dxdydz
+\int^\infty_{-\infty} \Psi_2^2 {{ \partial V_1} \over{
\partial(a/2)}}dxdydz. \label{hol23} \eeq After substituting $
\partial V_1/ \partial (a/2)=
\partial V_1/ \partial x$ the second
integral is done by parts and one arrives at: \beq {{
\partial\la\Psi_2|V_1|\Psi_2\ra} \over{ \partial(a/2)}}
=2\int^\infty_{-\infty} \Psi_2 \left[ {{ \partial\Psi_2} \over{
\partial(a/2)}} - {{ \partial\Psi_2} \over{ \partial x}} \right]
V_1dxdydz. \label{hol24} \eeq
>From the definition of $\Psi_2$ one
has:
\[
{{\partial \Psi_2} \over{ \partial(a/2)}} ={{\partial B} \over{
\partial(a/2)}}\phi_1 + {{\partial A} \over{ \partial(a/2)}}
\phi_2 - 2
\Gamma^2\left[B\left(x+a/2\right)\phi_1-A(x-a/2)\phi_2\right],
\]
\beq {{\partial \Psi_2} \over{ \partial x}}= -2 \Gamma^2
\left[B\left(x+a/2\right)\phi_1+A(x-a/2)\phi_2\right],
\label{hol24a} \eeq so that \beq {{ \partial\Psi_2} \over{
\partial(a/2)}} - {{ \partial\Psi_2} \over{ \partial x}}= {{
\partial B} \over{ \partial(a/2)}} \phi_1 +{{ \partial A} \over{
\partial(a/2)}} \phi_2 +4\Gamma^2A(x-a/2)\phi_2. \label{hol25} \eeq
Substituting \Eq{\ref{hol25}} into \Eq{\ref{hol24}}, developing
$\Psi_2=B\phi_1+A\phi_2$ and reordering yields finally:
\[
8\Gamma^2 \left ( AB\la\phi_1|xV_1|\phi_2\ra+A^2\la\phi_2|xV_1|\phi_2\ra
\right)=
{{\partial\la\Psi_2|V_1|\Psi_2\ra} \over{\partial(a/2)}}
-{{\partial B^2} \over{ \partial(a/2)}}
\la\phi_1|V_1|\phi_1\ra
\]
\beq +\left[4a\Gamma^2A^2- {{\partial A^2} \over{ \partial(a/2)}}
\right]\la\phi_2|V_1|\phi_2\ra + \left[4a\Gamma^2AB -2{{\partial
(AB)} \over{ \partial(a/2)}} \right]\la\phi_1|V_1|\phi_2\ra,
\label{hol26} \eeq which is the second equation needed to evaluate
the integrals entering the \Eq{\ref{hol6}} for $g_0^{(1)}$. For
the explicit evaluation we need also:
\[
\la\Psi_1|V_1|\Psi_1\ra=
2\Gamma\sqrt{2/\pi}\left[A^2+B^2F_0(2\Gamma^2a^2)+2ABSF_0(\Gamma^2a^2/2)\right],
\]
\beq \la\Psi_2|V_1|\Psi_2\ra=
2\Gamma\sqrt{2/\pi}\left[B^2+A^2F_0(2\Gamma^2a^2)
+2ABSF_0(\Gamma^2a^2/2)\right]. \eeq \noindent It is convenient to
subtract \ the \Eq{\ref{hol22}}  from \Eq{\ref{hol26}}  yielding:
\[
8\Gamma ^{2}\left( A^{2}-B^{2}\right) <\phi _{2}|xV_{1}|\phi
_{2}>=\frac{d} { d\left( a/2\right) }\left[ <\Psi _{2}|V_{1}|\Psi
_{2}>-<\Psi _{1}|V_{1}|\Psi _{1}>\right]
\]

\begin{equation}
+\frac{d\left( A^{2}-B^{2}\right) }{d\left( a/2\right) }<\phi
_{1}|V_{1}|\phi _{1}>+\left[ 4\Gamma ^{2}a\left( A^{2}-B^{2}\right) -\frac{%
d\left( A^{2}-B^{2}\right) }{d\left( a/2\right) }\right] <\phi
_{2}|V_{1}|\phi _{2}>, \label{cadd1}
\end{equation}
\noindent
while their sum yields:
\[
16\Gamma ^{2}AB<\phi _{1}|xV_{1}|\phi _{2}>+8\Gamma ^{2}\left(
A^{2}+B^{2}\right) <\phi _{2}|xV_{1}|\phi _{2}>=
\]

\[
\frac{d}{d\left( a/2\right) }\left[ <\Psi _{2}|V_{1}|\Psi _{2}>+<\Psi
_{1}|V_{1}|\Psi _{1}>\right] -\frac{d\left( A^{2}+B^{2}\right) }{d\left(
a/2\right) }<\phi _{1}|V_{1}|\phi _{1}>+
\]

\[
+\left[ 4\Gamma ^{2}a\left( A^{2}+B^{2}\right) -\frac{d\left(
A^{2}+B^{2}\right) }{d\left( a/2\right) }\right] <\phi _{2}|V_{1}{}|\phi
_{2}>
\]

\begin{equation}
+\left[ 8\Gamma ^{2}aAB-4\frac{d\left( AB\right) }{d\left(
a/2\right) }\right] <\phi _{1}|V_{1}{}|\phi _{2}>. \label{cadd2}
\end{equation}
\noindent To proceed, one has

\[
<\Psi _{2}|V_{1}|\Psi _{2}>-<\Psi _{1}|V_{1}|\Psi _{1}>=-\Gamma \sqrt{\frac{2%
}{\pi }}\left( A^{2}-B^{2}\right) \left[ 1-F_{0}\left( 2\Gamma
^{2}a^{2}\right) \right],
\]

\[
<\Psi _{2}|V_{1}|\Psi _{2}>+<\Psi _{1}|V_{1}|\Psi _{1}>=
\]

\begin{equation}
\Gamma \sqrt{\frac{2}{\pi }}\left\{\left( A^{2}+B^{2}\right)
\left[ 1+F_{0}\left( 2\Gamma ^{2}a^{2}\right) \right]
+4ABSF_{0}\left( \Gamma ^{2}a^{2}/2\right) \right\}. \label{cadd3}
\end{equation}
The derivatives of interest are:
\begin{equation}
\frac{\partial S}{\partial \left( a/2\right) }=-2\Gamma ^{2}aS,
\end{equation}

\begin{equation}
\frac{\partial \left( A^{2}+B^{2}\right) }{\partial \left( a/2\right) }=-%
\frac{4\Gamma ^{2}aS^{2}}{\left( 1-S^{2}\right) ^{2}}, \qquad
\frac{\partial \left( A^{2}-B^{2}\right) }{\partial \left( a/2\right) }=-%
\frac{2\Gamma ^{2}aS^{2}}{\left( 1-S^{2}\right) ^{3/2}}, \qquad
\frac{\partial AB}{\partial \left( a/2\right) }=\frac{\Gamma
^{2}aS\left( 1+S^{2}\right) }{\left( 1-S^{2}\right) ^{2}},
\label{cadd4}
\end{equation}

\begin{equation}
\frac{\partial }{\partial \left( a/2\right) }\left[ F_{0}\left(
2\Gamma ^{2}a^{2}\right) \right] =\frac{2}{a}\left[
S^{4}-F_{0}\left( 2\Gamma ^{2}a^{2}\right) \right], \qquad
\frac{\partial }{\partial \left( a/2\right) }\left[ F_{0}\left(
 {{\Gamma^{2}a^{2}}\over{2}}\right) \right] =
\frac{2}{a}\left[ S-F_{0}\left( {{\Gamma^{2}a^{2}}\over{2}}\right)
\right]. \label{cadd5}
\end{equation}
Therefore we obtain:

\[
\frac{\partial }{\partial \left( a/2\right) }\left[<\Psi _{2}|V_{1}|\Psi
_{2}>-<\Psi _{1}|V_{1}|\Psi _{1}>\right]=
\]

\begin{equation}
\left[ \frac{2\Gamma \sqrt{2/\pi }}{a\left( 1-S^{2}\right) ^{3/2}}%
\right] \left\{ \Gamma ^{2}a^{2}S^{2}\left[ 1-F_{0}\left( 2\Gamma
^{2}a^{2}\right) \right] +\left( 1-S^{2}\right) \left[
S^{4}-F_{0}\left( 2\Gamma ^{2}a^{2}\right) \right] \right\},
\label{cadd6}
\end{equation}
and

\[
\frac{\partial }{\partial \left( a/2\right) }[<\Psi _{2}|V_{1}|\Psi
_{2}>+<\Psi _{1}|V_{1}|\Psi _{1}>]=
\]

\[
\left[ \frac{2\Gamma \sqrt{2/\pi  }}{a\left( 1-S^{2}\right) ^{2}}%
\right] \{-2\Gamma ^{2}a^{2}S^{2}-\left( 2-S\right) S^{3}\left(
1-S^{2}\right) -F_{0}\left( 2\Gamma ^{2}a^{2}\right) \left[ 1-S^{2}\left(
1-2\Gamma ^{2}a^{2}\right) \right]
\]

\begin{equation}
+F_{0}\left( \frac{\Gamma ^{2}a^{2}}{2}\right) \left[ 4\Gamma
^{2}a^{2}S^{2}+2S^{2}\left( 1-S^{2}\right) \right] \}.
\label{cadd7}
\end{equation}
By substituting the equations above deduced and the other matrix
elements into the \Eq{\ref{cadd1}} and \Eq{\ref{cadd2}} we obtain
the unknown matrix elements as:

\beq \la\phi_2|\left(x-{a \over 2}\right)V_1|\phi_2\ra=
{{\sqrt{2/\pi} }
\over{2a\Gamma}}\left[S^4-F_0(2a^2\Gamma^2)\right], \label{hol27a}
\eeq \beq \la\phi_1|xV_1|\phi_2\ra= {{ S\sqrt{2/\pi} }
\over{a\Gamma }}\left[S
-F_0\left({{a^2\Gamma^2}\over{2}}\right)\right]. \label{hol27b}
\eeq
\section{Appendix 4}
Let us evaluate the crystal field coupling for site $1$ in the
adiabatic limit, in order to verify Eq.\ref{gcfaddef}.
For the charge on site 1 the variation of the local energy defines
$g_{cfad}^{(1)}$ as:
\[
\int^\infty_{-\infty}\Psi_1(\RIo-\uI,\RJo-\uJ,\rv)^2V_2(\RJo-\uJ,\rv)d^3\rv
-\int^\infty_{-\infty}\Psi_1(\RIo,\RJo,\rv)^2V_2(\RJo,\rv)d^3\rv
\]
\beq \equiv g_{cfad}^{(1)}n_1(u_2-u_1)+\cal{O} [(u_2-u_1)^2].
\label{cfad1} \eeq Substituting $t=x-a/2-u_2$ and developing
$\Psi_1(u,t,y,z)^2$ to first order in $u$ yields:

\beq g_{cfad}^{(1)}=2 \int^\infty_{-\infty} \Psi_1(0,t,y,z)
{{\partial \Psi_1} \over{\partial(u)}}\bigg\vert_{u=0} V_2(t,y,z)
\bullet \bfm{e}_{12}. \label{cfad4} \eeq By substituting
$t=x-a/2-u_2$ we obtain:
\[
{{\partial \Psi_1} \over{\partial u}}\bigg\vert_{u=0}=
{{\partial A} \over{\partial u}} \bigg\vert_{u=0}
Ne^{-\Gamma^2[(t+a)^2+y^2+z^2]}
+{{\partial B} \over{\partial u}} \bigg\vert_{u=0}
Ne^{-\Gamma^2[t^2+y^2+z^2]}
\]
\beq -2\Gamma^2A(t+a)Ne^{-\Gamma^2[(t+a)^2+y^2+z^2]}.
\label{cfad3} \eeq Substituting $\Psi_1$ and \Eq{\ref{cfad3}} into
\Eq{\ref{cfad4}}  yields:
\[
g_{cfad}^{(1)}=
{{\partial (A^2)} \over{\partial u}}\bigg\vert_{u=0}\la\phi_1|V_2|\phi_1\ra +
{{\partial (B^2)} \over{\partial u}}\bigg\vert_{u=0}\la\phi_2|V_2|\phi_2\ra +
2 {{\partial (AB)} \over{\partial u}}\bigg\vert_{u=0}\la\phi_1|V_2|\phi_2\ra
\]
\beq +4\Gamma^2\left[ A_0^2 \la\phi_1|(x+a/2)V_2|\phi_1\ra +
2A_0B_0\la\phi_1|xV_2|\phi_2\ra \right]. \label{cfad10} \eeq The
corresponding coupling for the other site is obtained by the site
permutation $1 \rightleftharpoons 2$, \beq
H_{cfad}^{(2)}=g_{cfad}^{(2)}\sum_\s n_{2\s}(u_1-u_2),
\label{cfad10b} \eeq that is, with $t=x+a/2-u_1$:

\beq g_{cfad}^{(2)} =2\int^\infty_{-\infty} \Psi_2(0,t,y,z)
{{\partial \Psi_2} \over{\partial(u_1-u_2)}}\bigg\vert_{0}
V_1(t,y,z). \label{cfad11} \eeq Performing the corresponding
calculations we obtain:
\[
g_{cfad}^{(2)} = g_{cfad}^{(1)} -4\Gamma^2\Biggl\{
AB \la\phi_1|x(V_1+V_2)|\phi_2\ra +A^2
(\left[ \la\phi_1|xV_2|\phi_1\ra +  \la\phi_2|xV_1|\phi_2\ra) \right]
\]
\beq +B^2 (\left[ \la\phi_2|xV_2|\phi_2\ra +
\la\phi_1|xV_1|\phi_1\ra) \right] \Biggr \}. \label{cfad12} \eeq
The difference between   $g_{cfad}^{(2)}$ and $ g_{cfad}^{(1)}$
vanishes because of the results reported in the Appendix 2, \Eq{
\ref{hol15}}.

\nonumsection{References}


\begin{thebibliography}{000}
\bibitem{acquarone1}We correct here a misprint in Ref.1.
In Eq.(1) there, the term $Vn_1n_2$ should read $(V-J_z/2)n_1n_2$.
The parameter $V$ in the present work corresponds therefore to
  $V-J_z/2$ there.
\end{thebibliography}
\end{document}